\documentclass[aps,prappl,twocolumn,groupedaddress,showkeys,showpacs,superscriptaddress,floatfix,longbibliography]{revtex4-1}
\usepackage{epsfig}
\usepackage{multirow}
\usepackage{amsmath, amssymb,mathtools}
\usepackage{color}
\usepackage{graphicx}
\usepackage{dcolumn}
\usepackage{bm}
\usepackage{subfigure}
\usepackage[utf8]{inputenc}
\usepackage[T1]{fontenc}
\usepackage{tikz}

\begin{document}

\title{Josephson-based threshold detector for L\'evy distributed current fluctuations}

\author{Claudio Guarcello\thanks{e-mail: claudio.guarcello@nano.cnr.it}}
\affiliation{NEST, Istituto Nanoscienze-CNR and Scuola Normale Superiore, Piazza S. Silvestro 12, I-56127 Pisa, Italy}
\affiliation{Centro de Fisica de Materiales (CFM-MPC), Centro Mixto CSIC-UPV/EHU, 20018 Donostia-San Sebastian, Basque Country, Spain}
\author{Davide Valenti\thanks{e-mail: }}
\affiliation{Dipartimento di Fisica e Chimica ``Emilio Segrè'', Group of Interdisciplinary Theoretical Physics, Università di Palermo and CNISM, Unità di Palermo, Viale delle Scienze, Edificio 18, 90128 Palermo, Italy}
\affiliation{Istituto di Biomedicina ed Immunologia Molecolare (IBIM) ``Alberto Monroy'', CNR, Via Ugo La Malfa 153, I-90146 Palermo, Italy}
\author{Bernardo Spagnolo\thanks{e-mail:}}
\affiliation{Dipartimento di Fisica e Chimica ``Emilio Segrè'', Group of Interdisciplinary Theoretical Physics, Università di Palermo and CNISM, Unità di Palermo, Viale delle Scienze, Edificio 18, 90128 Palermo, Italy}
\affiliation{Radiophysics Dept., Lobachevsky State University, 23 Gagarin Ave., 603950 Nizhniy Novgorod, Russia}
\affiliation{Istituto Nazionale di Fisica Nucleare, Sezione di Catania, Via S. Sofia 64, I-95123 Catania, Italy}
\author{Vincenzo Pierro\thanks{e-mail: }}
\affiliation{Dept. of Engineering, University of Sannio, Corso Garibaldi 107, I-82100 Benevento, Italy}
\affiliation{INFN, Sezione di Napoli Gruppo Collegato di Salerno, Complesso Universitario di Monte S. Angelo, I-80126 Napoli, Italy}
\author{Giovanni Filatrella\thanks{e-mail: }}
\affiliation{Dep. of Sciences and Technologies and Salerno unit of CNISM, University of Sannio, Via Port’Arsa 11, Benevento I-82100, Italy}

\date{\today}

\begin{abstract}
We propose a threshold detector for L\'evy distributed fluctuations based on a Josephson junction. 
The L\'evy noise current added to a linearly ramped bias current results in clear changes in the distribution of switching currents out of the zero-voltage state of the junction. 
We observe that the analysis of the cumulative distribution function of the switching currents supplies information on both the characteristics shape parameter $\alpha$ of the L\'evy statistics and the intensity of the fluctuations. 
Moreover, we discuss a theoretical model which allows to extract characteristic features of the L\'evy fluctuations from a measured distribution of switching currents. In view of this results, this system can effectively find an application as a detector for a L\'evy signal embedded in a noisy background.
\end{abstract}

\pacs{}

\maketitle

\section{Introduction}
\label{Intro}
\begin{figure}[b!!]
\centering
\includegraphics[width=\columnwidth]{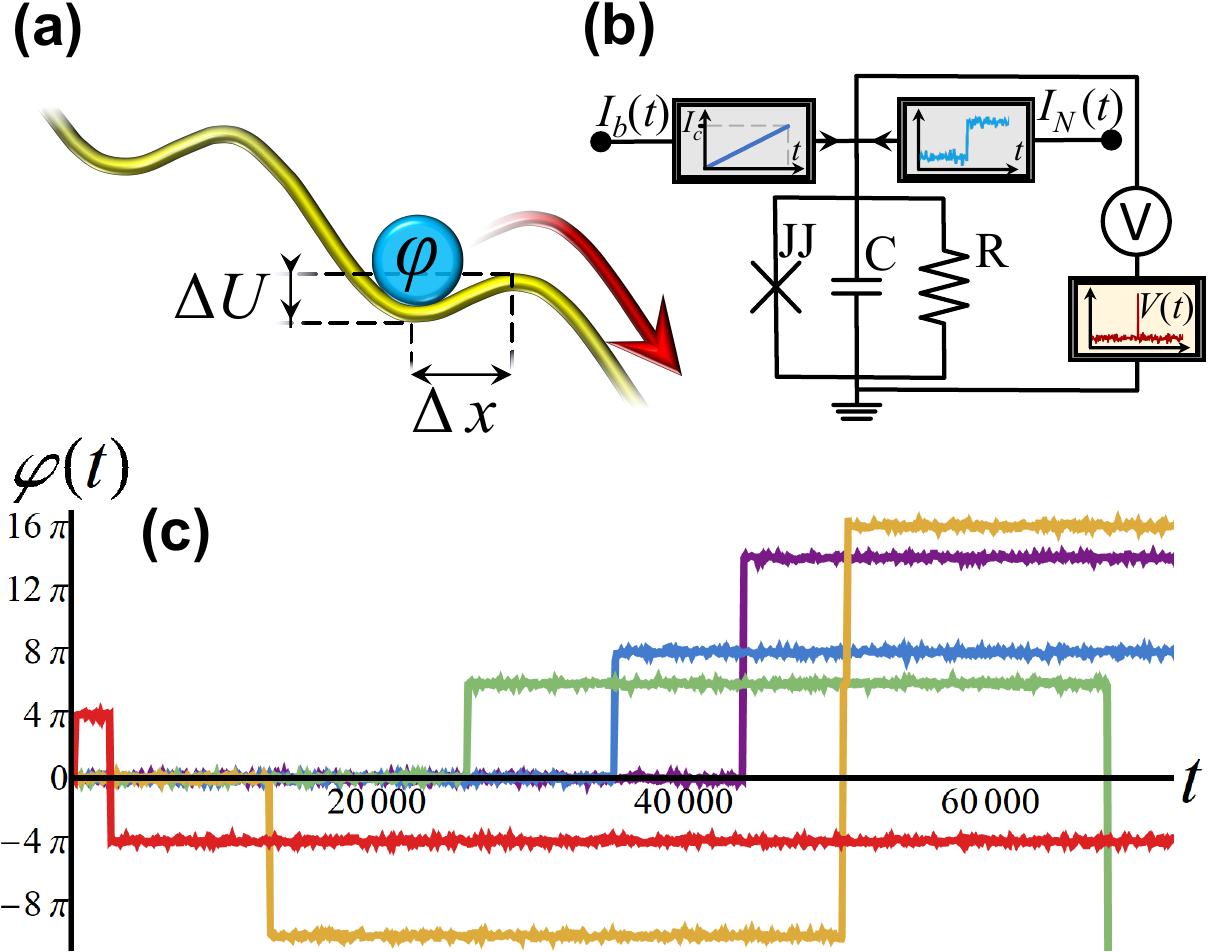}
\caption{
(a) The phase particle in a potential minimum of the tilted washboard potential $U$. The barrier height, $\Delta U$, and the distance between the minimum and maximum of the potential, $\Delta x$, are also shown. 
(b) $\varphi$ trajectories in the noise driven case when the L\'evy stochastic term dominates.
(c) Simplified equivalent circuit diagram for the resistively and capacitively shunted junction model. The linearly ramped bias current, $I_b(t)$, and the noise current, $I_N(t)$, of the JJ are included in the diagram.}
\label{Fig01}
\end{figure}

\vskip-0.2cm
A current-biased Josephson junction (JJ) represents a natural threshold detector for current fluctuations, inasmuch as it is a metastable system operating on an activation mechanism. 
In fact, the behaviour of a JJ can be depicted as a particle, representing the superconducting phase difference $\varphi$ across the JJ, in a cosine ``washboard'' potential with friction~\cite{Bar82,Lik86}, see Fig.~\ref{Fig01}(a).
In this picture, the slope of the washboard potential is given by the injected current, and the dynamics of the phase is described by the resistively and capacitively shunted junction (RCSJ) model. 
The equivalent particle remains near a washboard minimum (correspondingly, the JJ is in the so-called zero-voltage metastable state) until the direct bias current exceeds a critical value, or a fluctuation sets the phase $\varphi$ in motion along the potential. 
In fact, a current fluctuation instantaneously tilts the potential, so that a noise-induced escape from a minimum can occur. 
In correspondence of the escape a voltage develops, as the voltage is related to the velocity of the phase particle.
Shortly, if a JJ is set in the fundamental zero-voltage state, the noise can cause a passage from this zero-voltage state to the finite voltage ``running'' state.
The statistics of these passages can be exploited to reveal the features of the noise.

In this work we address the issue of the characterization of a specific kind of non-Gaussian fluctuations, namely, the $\alpha$-stable L\'evy noise, through the switching currents distribution of a JJ. These stochastic processes drive the virtual particle, namely, the phase $\varphi$ in the Josephson context, over a very long {\it distance} in a single displacement, namely, a {\it flight}. To visualize the effects produced by a L\'evy noise source on the JJ's behavior, in Fig.~\ref{Fig01}(b) we show several $\varphi$ trajectories, obtained in the noise driven case and in the absence of bias current, that are characterized by abrupt fluctuations. 

Results on the dynamics of systems driven by L\'evy flights have been recently reviewed in Refs.~\cite{Dub08, Zab15}. 
L\'evy flights well describe transport phenomena in different condensed matter contexts and practical applications. 
For instance, in graphene the presence of L\'evy distributed fluctuations has been recently discussed~\cite{Bri14,Gat16}. Specifically, graphene stripe with anisotropically distributed on-site impurities shows L\'evy flight transport in the stripe direction~\cite{Gat16}, and it has also been proposed that the particular electron-electron interaction of the graphene electronics can produce a L\'evy flights distribution as a response to a laser source~\cite{Bri14}. Moreover, it has been speculated that the anomalous premature switches affecting the switching currents in graphene-based JJs, that are likely to be unrelated to thermal fluctuations~\cite{Cos12}, could be ascribed to L\'evy distributed phenomena, see Ref~\cite{Gua17} where the nonsinusoidal potential appropriated for graphene JJs \cite{GuaVal15,convegno,Spa17} has been investigated. Consequently, the response of any graphene-based device could be intrinsically affected by L\'evy distributed fluctuations.

Still dealing with material issues, also photoluminescence experiments in moderately doped $n$-InP samples reveal anomalous L\'evy-type distribution~\cite{Lur10,Lur12,LurSub12,Sem12,Sub13,Sub14}. This phenomenon could have a large impact on the design of a number of optoelectronic devices such as multicolored LEDs, opto-thyristors, photovoltaic devices with high efficiency, or semiconductor scintillator for radiation detection~\cite{Sub13}. 
Moreover, L\'evy processes emerge also in the electron transport~\cite{Nov05} and optical properties~\cite{Kun00,Kun01,Shi01,Mes01,Bro03} of semiconducting nanocrystals quantum dots.

Looking instead at thermal properties of materials, it has been shown that the quasiballistic heat conduction in semiconductor alloys is governed by L\'evy superdiffusion~\cite{Ver15I,Ver15II,Moh15,Upa16}. The L\'evy engineering of heat can also impact the thermal conductivity and may offer novel ideas towards thermal conductivity reduction for thermoelectric applications.

On a more applicative side, L\'evy noise often appear in telecommunications and networks~\cite{Yan03,Bha06,Cor10}. In fact, in some communication channels, noise exhibits impulsive, L\'evy-type, as well as Gaussian, characteristics. The source of impulsive noise may be either natural or man-made, it may include atmospheric noise or ambient noise, and it might come from relay contacts, electromagnetic devices, electronic apparatus, or transportation systems, switching transients, and accidental hits in telephone lines~\cite{Bha06}. The demand to recognize these disturbances led to several proposals for models and detection schemes in this framework~\cite{Tsi95,Sub15,Sho15,Vin15}.

Finally, concerning a purely engineering issue, L\'evy fluctuations have been also used to describe vibration data in industrial bearings~\cite{LiYu10,Cho14,Saa15} and in wind turbines rotation parts~\cite{Ely16}. In fact, when a rolling element bearing runs in fault condition, the observed vibration signal from the bearing is a non-Gaussian signal with impulsive behavior~\cite{Whi84,McF84}. A detection tool for these signals provides a direction for rotating machine fault diagnosis.

The cases discussed so far demonstrate the importance of a reliable tool capable of detecting fluctuations distributed according to L\'evy statistics. We demonstrate that a noise detector based on JJ is suitable for studying this type of fluctuations.

Nowadays, following the seminal suggestions of Refs.~\cite{Tob04,Pek04,Ank05}, several experimental setups of Josephson-based noise detectors have been realized~\cite{Pek05,Ank07,Suk07,Tim07,Pel07,Hua07,Gra08,LeM09,Urb09,Fil10,Add12,Oel13,Add13,Sol15,Sai16}. 
A scheme to detect the Poissonian character of the charge injection in an underdamped JJ, based on the analysis of the third-order moment of the electrical noise, was proposed in Ref.~\cite{Pek04} and a scheme to detect the fourth-order moment of the noise has been analyzed in Ref.~\cite{Ank05}. 
A threshold detector based on an array of overdamped JJs for the direct measurement of the full counting statistics, through rare over-the-barrier jumps induced by current fluctuations, was suggested in Ref.~\cite{Tob04}. Alternatively, in the Coulomb blockade regime, the sensitivity of the JJ conductance to the non-Gaussian character of the applied noise was demonstrated~\cite{Lin04,Hei04}. Most proposals make use of the information content of higher moments, beyond the variance, of the electric noise, mainly to discuss the Poissonian character of the current fluctuations. 
However, the deviations from the Gaussian behaviour are typically small and experimental measurements of third and fourth moments are actually demanding and error-prone with respect to measurements of dc-transport properties. 
Moreover, a L\'evy flights distribution exhibits power-law tails and, consequently, second and higher moments diverge. 
This feature eventually poses a relevant complication in relating L\'evy flight models to experimental data, since the latter, due to the infinite variance of the noise source, can suffer limitless-intensity fluctuations. 
A JJ-based threshold detector could circumvent this difficulty, since the switching occurs as the phase particle passes a potential barrier, regardless the intensity of fluctuations. 
The distribution of the current values in correspondence of which a switch occurs, i.e., the switching currents $i_{SW}$, catches the information content we are interested in. 
Therefore, the investigation of JJ switching currents could pave the way for the direct experimental investigation of an $\alpha$-stable L\'evy noise signal or the L\'evy component of an unknown noise signal.

In this work, we address the problem of the experimental estimation of the parameters of the L\'evy noise source in the context of Josephson-based detectors. Accordingly, we first describe the physical phenomenon behind the problem, namely, how the L\'evy noise affects the switching towards the resistive state of a JJ biased by a slowly, linearly increasing electric current. Then, we study an effective way to deduce the information we are interested in from  available experimentally data, namely, the cumulative distribution functions (CDFs) of the switching currents. Therefore, we propose to employ a well-established device, namely, a Josephson junction, in the context of the L\'evy noise detection. We also argue that the CDF of the switching currents is a convenient quantity for such a detection. Finally, the theoretical simulations that we perform are supported by analytical estimates.

The paper is organized as follows. In Sec.~\ref{Model}, the theoretical background used to describe the phase evolution of a short JJ is discussed. Moreover, both the statistical properties of the L\'evy noise and the power-law asymptotic behaviour of the mean escape time are briefly reviewed. In Sec.~\ref{Results}, the theoretical results are shown and analyzed. We give also an analytical estimate of the distributions of switching currents in the presence of a L\'evy noise source. In Sec.~\ref{Conclusions}, conclusions are drawn.

\section{Model}
\label{Model}\vskip-0.2cm

A typical setup for a Josephson based noise readout~\cite{Pek05,Suk07,Urb09} consists of a JJ on which two superimposed currents, $I_b(t)$ and $I_N(t)$, are flowing [see Fig.~\ref{Fig01}(c)]. Specifically, $I_b(t)$ is the deterministic bias current drawn from a parallel source and $I_N(t)$ is the stochastic current.
We neglect escapes guided by macroscopic quantum tunnelling~\cite{Gra84} to consider exclusively processes activated by thermal as well as non-Gaussian fluctuations.

A measurement consists in slowly and linearly ramping the bias current in a time $t_{\text{max}}$, so that $I_b(t_{\text{max}})=I_c$ ($I_c$ is the critical current of the JJ), and to record the value at which a switch occurs.
In this readout scheme, the noise influence is considered in the limit of adiabatic bias regime, where the change of the slope of the potential induced by the bias current is slow enough to keep the phase particle in the metastable well until the noise pushes out the particle. 
Finally, after the time $t_{\text{max}}$, a ``reset'' is performed driving the bias current down to zero. 
In this work, sequences of $10^4$ ramps of maximum duration $t_{\text{max}}=10^7\omega_p^{-1}$ are applied to the junction, where $\omega_{p}=\sqrt{2eI_c/(\hbar C)}$ and $C$ are the plasma frequency and the capacitance of the JJ, respectively. 
Finally, a distribution of switching currents is obtained.

The phase dynamics is obtained numerically solving the RCSJ model equation~\cite{Bar82}
\begin{equation}
\left ( \frac{\Phi_0}{2\pi} \right )^2 C \frac{d^2 \varphi}{d t^2}+\left ( \frac{\Phi_0}{2\pi} \right )^2\frac{1}{R} \frac{d \varphi}{d t}+\frac{d }{d \varphi}U = \left ( \frac{\Phi_0}{2\pi} \right ) I_N,
\label{RCSJ}
\end{equation}
where $\Phi _0=h/(2e)\simeq 2.067\times 10^{-15}\textup{ Wb}$ is the flux quantum and $R$ is the normal resistance of the JJ. Here, $U$ is the washboard potential [see Fig.~\ref{Fig01}(a)]
\begin{equation}
U=U_0\left [1- \cos(\varphi) -i_b\varphi\right ],
\label{Washboard}
\end{equation}
where $U_0=\left ( \Phi_0/2\pi \right )I_c$. 
The average slope of the potential $U$ is given by the normalized ramped bias current, $i_b(t)=I_b(t)/I_c=v_bt$, where $v_b=t_{\text{max}}^{-1}$ is the ramp speed. The resulting activation energy barrier $\Delta U=2\left [ \sqrt{1-i_b^2} -i_b\arcsin(i_b)\right ]$ confines the phase $\varphi$ in a potential minimum. 

Eq.~\eqref{RCSJ} can be recast for convenience in a compact form
\begin{equation}
m \frac{d^2 \varphi}{d t^2} + m \eta \frac{d \varphi}{d t} + U_0 \frac{d }{d \varphi}u = U_0 i_N,
\label{RCSJ_c}
\end{equation}
where $m=\left ( \Phi_0/2\pi \right )^2C$ is the effective junction mass, the friction is governed by the parameter $\eta = 1/(RC)$, $u=U/U_0$, and $i_N=I_N/I_c$ is the stochastic term. 
In these units, $\omega_p=\sqrt{U_0/m}$.
Thermal noise is assumed to be negligible with respect to the L\'evy noise, that is, the JJ is cooled to temperatures where the Johnson-Nyquist contribution can be neglected.
In all simulations we assume the damping $\eta = 0.1 \omega_p$, the ramp speed $v_b =10^{-7}\omega_p$, and the L\'evy noise intensity $D= 5\times 10^{-7}$.

To model the L\'evy noise sources, we use the algorithm proposed by Weron~\cite{Wer96} for the implementation of the Chambers method~\cite{Cha76}. 
The notation $S_{\alpha}(\sigma, \beta, \lambda)$ is used for the L\'evy distributions~\cite{Dub07,Gua13,Val14,Spa15,Gua15,Gua16}, where $\alpha\in(0,2]$ is the {\it stability index}, $\beta\in[-1,1]$ is called {\it asymmetry parameter}, and $\sigma>0$ and $\lambda$ are a scale and a location parameter, respectively. The stability index characterizes the asymptotic long-tail power law for the distribution, which for $\alpha<2$ is of the $\left | x \right |^{-\left ( 1+\alpha \right )}$ type, while for $\alpha=2$ is the Gaussian distribution. 
We consider exclusively symmetric (i.e., with $\beta=0$), bell-shaped, standard (i.e., with $\sigma=1$ and $\lambda=0$), stable distributions $S_{\alpha}(1, 0, 0)$, with $\alpha\in(0,2)$.

{\it L\'evy escape. -- }
Escapes over a barrier in the presence of L\'evy noise have been thoroughly investigated for the overdamped case~\cite{Dyb06,Dyb07,Dub08,Dub09}. 
If both the distance between neighbour minimum and maximum of a metastable potential and the height of the potential barrier [see Fig.~\ref{Fig01}(a)] are unitary ($\Delta x = 1$ and $\Delta U = 1$, respectively), the power-law asymptotic behaviour of the mean escape time $\tau$ for the L\'evy statistics reads \cite{Che05,Che07}
\begin{equation}
\tau\left ( \alpha,D \right )=\frac{{\cal C}_{\alpha}} {D^{\mu_{\alpha}} },
\label{tau_Sw}
\end{equation}
where both the power-law exponent $\mu_\alpha$ and the coefficient ${\cal C}_\alpha$ depend on $\alpha$.
For arbitrary spatial and energy scale, by rescaling time, energy, and space in the overdamped case of Eq.~\eqref{RCSJ_c}~\cite{Che05}, Eq.~\eqref{tau_Sw} is replaced by
\begin{equation}
\tau\left ( \alpha,D \right ) = 
\left (\frac{ \eta ^{1-\mu_{\alpha}} \Delta x^{2-2\mu_{\alpha} + \alpha \mu_{\alpha}} } 
 {4^{1-\mu_{\alpha}} \Delta U^{1-\mu_{\alpha}} 2^{\alpha \mu_{\alpha}}} \right )
\frac{{\cal C}_{\alpha}} 
{D^{\mu_{\alpha} }} .
\label{tau_Levy_gen}
\end{equation}
The scaling exponent $\mu_{\alpha}$ and the coefficient ${\cal C}_\alpha$ are supposed to have a universal behaviour for overdamped systems, in particular $\mu_{\alpha} \simeq 1+ 0.0401 \left(\alpha - 1\right) + 0.105 \left(\alpha - 1\right)^2$~\cite{Che05}.
Then, by assuming $\mu_{\alpha} \simeq 1$ in the prefactor, Eq.~\eqref{tau_Levy_gen} becomes~\cite{Che05,Che07,Dub08}
\begin{equation}
\tau\left ( \alpha,D \right ) = 
\left ( \frac{ \Delta x } {2} \right )^{\alpha} 
\frac{{\cal C}_{ \alpha}} {D^{\mu_{\alpha} }}.
\label{tau_Levy}
\end{equation}
The physical interpretation of the previous assumption is that in the presence of L\'evy flights the mean escape time is independent on the barrier height $\Delta U$ and only depends upon the distance $\Delta x$.
The above equation is analogous to the Kramers rate for Gaussian noise~\cite{Kra40}.
The escape rate $\tau$ is inversely proportional through the coefficient $\mathcal{C}_{\alpha}$ (see Eq.~\eqref{tau_Levy}) to the noise parameters $D$.

Eqs.~\eqref{tau_Sw},~\eqref{tau_Levy_gen}, and~\eqref{tau_Levy}, are obtained and strictly valid only in the overdamped regime, i.e., $\eta/\omega_p \gg 1$. 
We speculate that the formula still holds for the moderately underdamped case, i.e., for $0.1 \le \eta/\omega_p \le 1$.

\begin{figure}[t!!]
\centering
\includegraphics[width=\columnwidth]{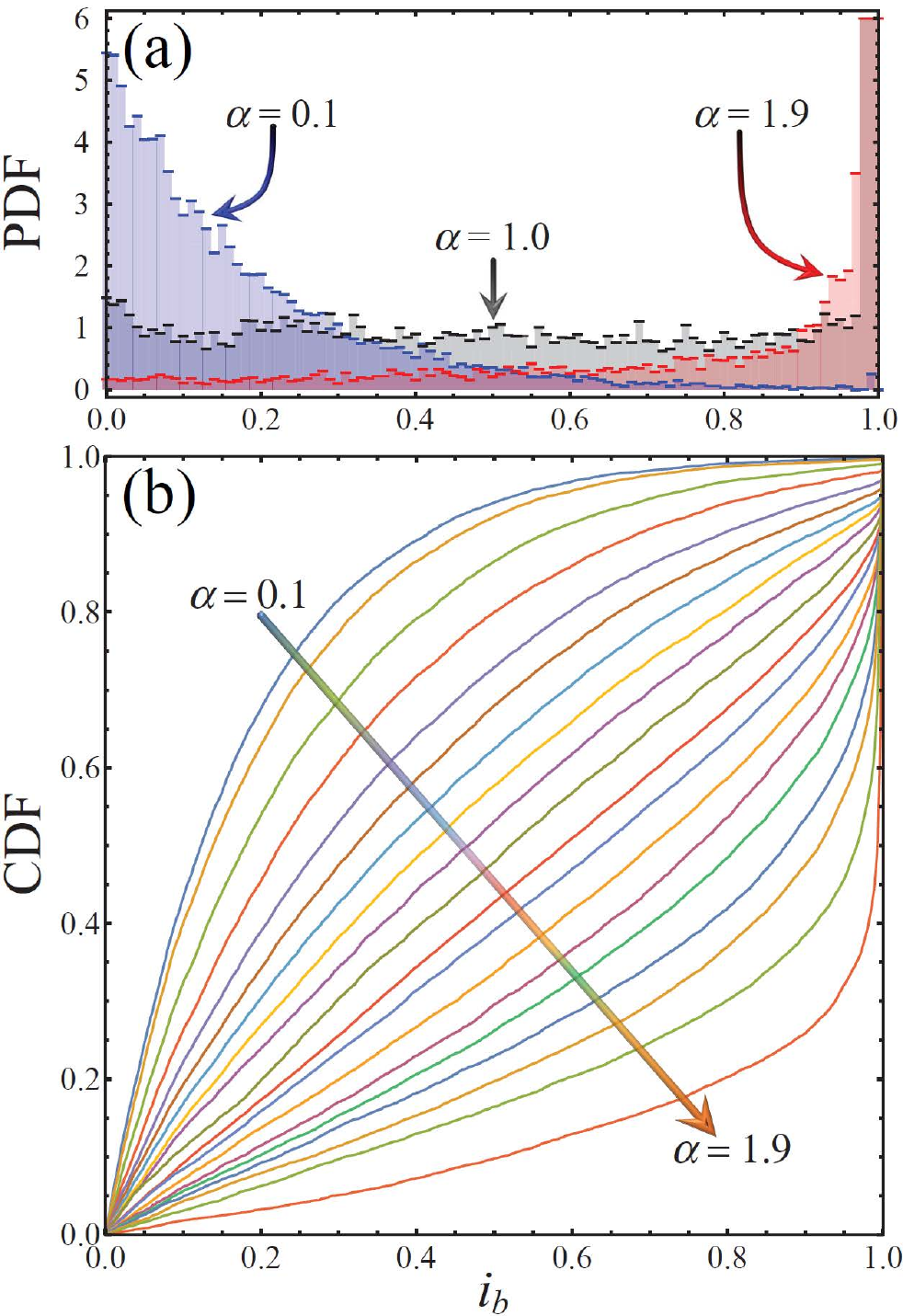}
\caption{
(a) Probability distribution function of the switching currents $i_{SW}$ for three cases of L\'evy fluctuations: $\alpha=0.1$ (leftmost peaked data), $\alpha = 1$ (flat data), and $\alpha = 1.9$ (rightmost peaked data).
(b) Cumulative distribution function of $i_{SW}$ for $\alpha$ values in the range $0.1 \div 1.9$. Parameters of the simulations are: $D= 5\times 10^{-7}$, $v_b =10^{-7}\omega_p$, and $\eta = 0.1 \omega_p$.}
\label{Fig02}
\end{figure}

\section{Results}
\label{Results}\vskip-0.2cm

The switching currents $i_{SW}$, are the experimental evidence of the escape processes in JJs. 
A collection of escapes can be characterized by a probability distribution function (PDF) of switching currents, as shown in Fig.~\ref{Fig02}(a) for three peculiar cases, $\alpha=0.1, 1.0, 1.9$. 
For the lowest $\alpha$ value, i.e., $\alpha=0.1$, the PDF resembles an exponential distribution. For $\alpha=1$, i.e., the Cauchy-Lorentz distribution, the PDF is roughly flat. Finally, for $\alpha\to2$, the distribution approaches the PDF of a Gaussian noise.
The most evident distinction between L\'evy and Gaussian cases lies in the low currents behaviour of the PDFs: in the former case, the switching probability is sizable, while in the latter case it is vanishingly small~\cite{Ful74}.

The effect of the parameter $\alpha$ is further elucidated in the cumulative distribution functions (CDFs), namely, the probability that $i_{SW}$ takes a value less than or equal to the bias current $i_b$, shown in Fig.~\ref{Fig02}(b). Here, we note that each CDF at a given value of $i_b$ decreases with $\alpha$. Therefore, the CDFs are suitable for the estimation of $\alpha$~\cite{Gua16}.
To model the CDFs of the switching currents, we exploit Eq.~\eqref{tau_Levy} to describe the escape rates over a barrier.
The average escape times estimated by Eq.~\eqref{tau_Levy} allow to connect the switching currents with the properties of the L\'evy noise. 
The CDF of $i_{SW}$ as a function of $i_b$ for a specific initial value of the bias ramp, $i_0$, reads
\begin{equation}
\text{CDF}(i_b |i_0)=1- \mathrm{Prob} \left[ i _{SW} > i_b |i_0 \right ].
\label{eq:CDF_ib}
\end{equation}

Recalling that the distribution of the escape times is exponential with rate $1/\tau(i_b)$ also for L\'evy flight noise~\cite{Che07}, the same logic of the seminal paper \cite{Ful74} leads to the expression 
\begin{equation}
P(i_{b}|i_0) = \mathcal{N}~ \frac{1}{v_b}~ \frac{1}{\tau(i_b)} \exp\left [ -\frac{1}{v_b} \int_{i_0}^{i_b} \frac{1}{\tau(i) } d i \right ]
\label{Pib}
\end{equation}
for the PDF associated to Eq.~\eqref{eq:CDF_ib} as a function of the average escape time $\tau (i_b)$. Here, $ \mathcal{N}$ is an appropriated normalization constant.
Eq.~\eqref{Pib} makes it evident the dependence of the switching current distribution $P(i_{b}|i_0) $ on the average escape time, $\tau (i_b)$, which is related to the noise features. 
For the thermal noise, Kramers' formula entails that the escapes across the barrier depend on the barrier height. 
For L\'evy noise, with the same widely employed approximations behind Eq.~\eqref{tau_Levy}, $\tau(i_b)$ turns out independent of the barrier height $\Delta U$, and becomes only function of $\Delta x$, which in turn via Eq.~\eqref{Washboard} depends on $i_b$ through the relation $\Delta x = \pi -2\arcsin{i_b}$.
The expression of $\tau(\alpha,D)$, Eq.~\eqref{tau_Levy}, inserted in Eq.~\eqref{Pib} gives for the L\'evy statistics (at the first order in $i_b$)
\begin{equation}
P(i_b|i_0) \propto \exp \left [ 
- \left( \frac {2} { \pi } \right)^{\alpha} 
\frac{i_b D^{\mu_{\alpha}} } {{\cal C}_{\alpha}v_b} \right ].
\label{P_t}
\end{equation}
Since the above equation contains the explicit expression for the argument of the exponential, it is a further step forward with respect to the results of Ref.~\cite{Gua17}. This breakthrough paves the way towards the applications of a Josephson junction as a L\'evy noise detector.

Notably, the solution of Eq.~\eqref{Pib} can be analytically computed and expressed in a compact form by using the nonlinear function ${\cal F}_{\alpha}$ defined as
\begin{eqnarray}
&{\cal F}_{\alpha} (i_b) = 2^{\alpha } \left\{\frac{\cosh^{-1}\left(i_b\right)}{2\left[\pi -2 \arcsin \left(i_b\right)\right]^{\alpha }} \Big [E_{\alpha }\left(\cosh^{-1}\left(i_b\right)\right)+\;\;\qquad\right. \\ 
&\left.-E_{\alpha }\left(-\cosh^{-1}\left(i_b\right)\right)\Big ] + \frac{i \pi ^{1-\alpha } }{4} \Big [E_{\alpha }\left(-\frac{i \pi }{2}\right)-E_{\alpha }\left(\frac{i \pi }{2}\right)\Big ]\right\},\nonumber
 \label{Faux}
 \end{eqnarray}
where $E_{\alpha}$ is the exponential integral with $\alpha$ argument~\cite{Prudnikov98}. 
Then, the PDF can be written as
\begin{eqnarray}
P(i_b|i_0) = {\cal N} 
\frac{\mathrm{d} {\cal F}_{\alpha}}{\mathrm{d} i_b}
\exp \left\{ -\frac{D^{\mu_{\alpha}}}{{\cal C}_{\alpha} v_{b}} \Big [ {\cal F}_{\alpha}(i_b)-{\cal F}_{\alpha}(i_0) \Big ] \right\},\qquad
\label{PDFib}
\end{eqnarray}
where the normalizing factor ${\cal N}$ reads
\begin{equation}
{\cal N} = \left\{ 1 - \exp{ \left[ -\frac{D^{\mu_{\alpha}}}
{{\cal C}_{\alpha} v_{b}} \Big ( {\cal F}_{\alpha}(1)-{\cal F}_{\alpha}(i_0) \Big ) \right] } \right\}^{-1} .
\end{equation}
The corresponding CDF is
\begin{equation}
\text{CDF}(i_b|i_0)= {\cal N} \left\{ 1 - \exp \left[ -\frac{D^{\mu_{\alpha}}}{{\cal C}_{\alpha} v_{b}} \Big ( {\cal F}_{\alpha}(i_b)-{\cal F}_{\alpha}(i_0) \Big ) \right] \right\}.
\label{CDFib}
\end{equation}

\begin{figure}[t!!]
\centering
\includegraphics[width=\columnwidth]{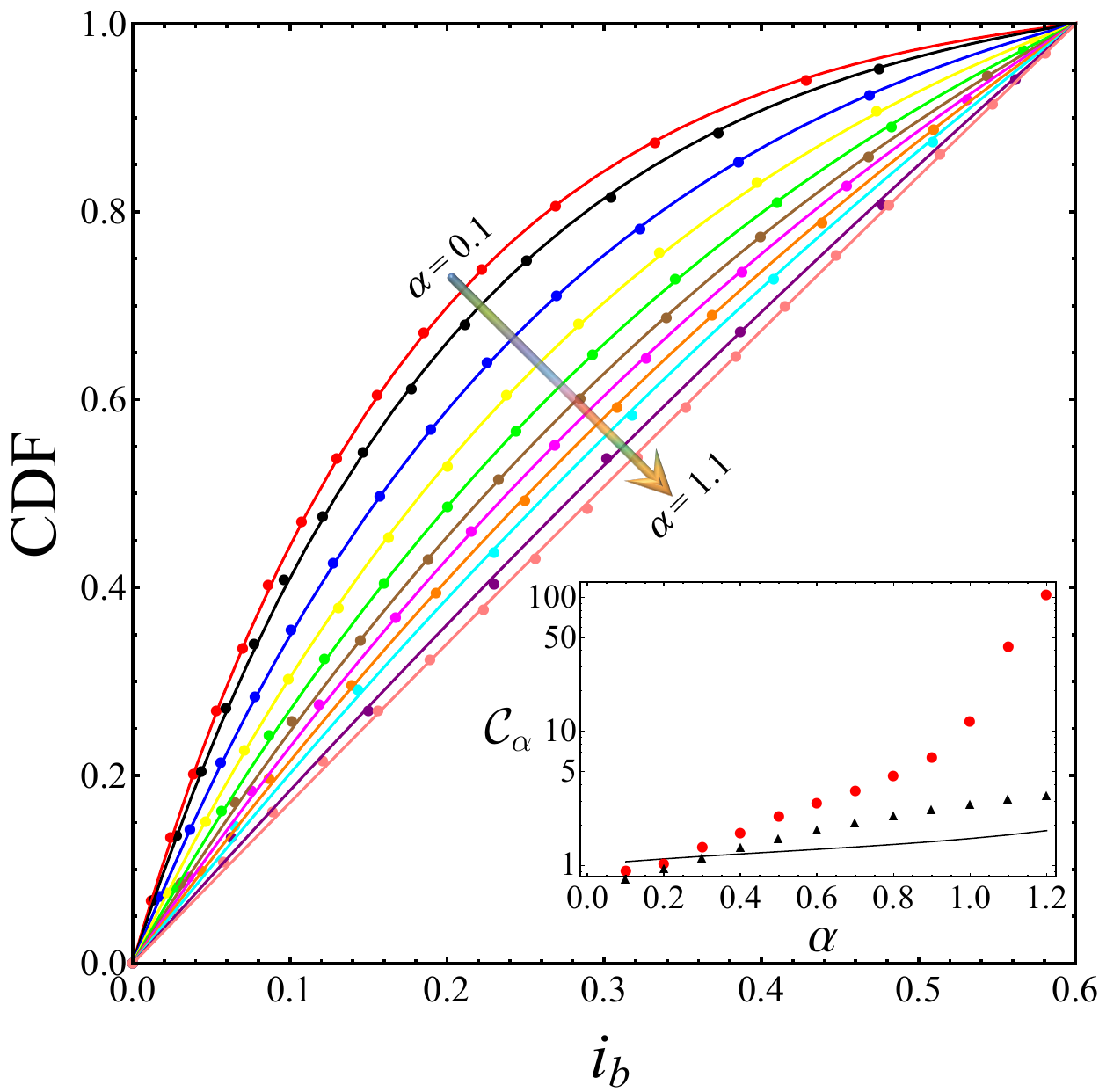}
\caption{
Marginal, i.e., obtained for $i_b \leq 0.6$, L\'evy noise induced CDFs of $i_{SW}$ computed by numerical solution of Eq.~\eqref{RCSJ_c} (solid lines) for $\alpha =0.1 \div 1.1$, $D=5\times 10^{-7}$, $v_b=10^{-7}\omega_p$, and $\eta = 0.1 \omega_p$. 
The theoretical curves obtained by numerical fitting of Eq.~\eqref{CDFib} are also reported for comparison (full circles).
In the inset we show the estimate of the coefficient ${\cal C}_{\alpha}$ based on the numerical fitting of Eq.~\eqref{CDFib} of the marginal CDFs displayed in the main panel (red circles). 
For comparison, we show also the numerical estimates given in Ref.~\cite{Che07} (black triangles) and the analytical estimate of Ref.~\cite{Che05}, namely, $\Gamma (1-\alpha) \cos(\pi \alpha/2 )$ (solid line). }
\label{Fig03}
\end{figure}

This is the main result of this work, that is to connect the properties of L\'evy flights with the accessible quantity of the switching currents distribution.
It is important to remind the main approximations underlying Eq.~\eqref{CDFib}: it has been assumed that the result obtained for an overdamped system, see Eq.~\eqref{tau_Levy}, still holds for moderately underdamped systems, and that Eq.~\eqref{Pib} can be applied to a slowly varying process ruled by the L\'evy escape time, see Eq.~\eqref{tau_Levy}.

We have performed extensive numerical simulations to check the validity of the results, given by Eqs.~\eqref{tau_Levy} and~\eqref{CDFib}. 
In the main panel of Fig.~\ref{Fig03} we show the marginal CDF, i.e., restricted to the maximum bias value $i_b = 0.6$, for $\alpha = 0.1 \div1.1$, $D = 5\times10^{-7}$, $v_b = 10^{-7}\omega_p$, and $\mu_\alpha = 1$.
The choice of these values for $\alpha$ and $i_b$ arises from practical considerations, since Eqs.~\eqref{tau_Levy} and~\eqref{CDFib} are more accurate for low bias currents and low $\alpha$ values, respectively. 
For these values the L\'evy flight jump features dominate, while in the opposite limits, $i_b \simeq 1$ and $\alpha \simeq 2$, the Gaussian characteristics set in. 
Accordingly, in the considered range of values, the effects of the Gaussian noise contribute can be safely ignored.
The numerical curves obtained by fitting of Eq.~\eqref{CDFib} are also reported for comparison in the main panel of Fig.~\ref{Fig03}.
The agreement between the computational results and the theoretical analysis, see Eq.~\eqref{CDFib}, is quite accurate for $\alpha < 1$. For $\alpha \gtrsim 1$ the statistics of the switches becomes undistinguishable from the uniform distribution (the bisector in Fig. 3). Thus, the model we proposed can be used to determine the value of $\alpha$ from switching currents measurements (as the other parameters are known), but it proves to be especially valuable in the region $\alpha < 1$.

In the inset of Fig. 3 we show with red circles the estimate of the coefficient $C_{\alpha}$ obtained by numerical fitting of Eq.~\eqref{CDFib} of the marginal CDFs shown in the main panel. The estimates of the values of $C_{\alpha} \gtrsim 1$ significantly deviate from both the numerical estimates given in Ref.~\cite{Che07} and the analytical estimate obtained in Ref.~\cite{Che05}. However, these differences can be ascribed to: \emph{i}) an overdamped rather than underdamped dynamics; \emph{ii}) a fixed rather than a slowly varying potential barrier; \emph{iii}) a cubic rather than a cosine potential.\\

The numerical simulations required to design the detector proposed in this work are realistic, even if they take a long time, on standard processor nowadays available, with respect to a real experimental realization. However, it is possible to keep the simulations requirements at bay and still reproduce realistic values of the parameters. To estimate the simulation time, one can set the bias ramp time $t_{\text{max}}=10^7\omega_p^{-1}\approx0.01\;\text{ms}$ (if one takes the plasma frequency $\omega_p$ of the order of one THz). This time guarantees an adiabatic evolution, even if it is definitively shorter than the realization time of a real experimental switching current measurement, c.f., Ref.~\cite{Cos12}. To build a distribution of switching currents, we perform a sequence of $10^4$ numerical realizations. Thus, the total simulation time for a single distribution is still feasible, although very long. Nevertheless, stochastic simulations can be extensively parallelized~\cite{Pie18} taking advantage of fast and low cost general purpose graphic process units. The estimation of the minimum number of realizations required to detect the presence of a L\'evy noise component can be done through a Kolmogorov-Smirnov (KS) test (see Ref.~\cite{Gua17}). We choose the KS test, because it is the optimal detection technique to discriminate between two different CDFs, in our case between L\'evy flights and Gaussian noise induced switches.
It can be shown that with a number of $10^3$ trials it is possible to achieve a $p$-value below 1$\%$~\cite{Gua17}. Furthermore, it can be shown that the higher the noise level, the lower the number of data necessary
to confirm the presence of the L\'evy noise component.
To estimate the measurement time to obtain a single distribution of the switching currents, one can start from the observation that typical experimental measurements of switching current distributions consist of about $N_{\text{exp}}\lesssim10^4$~\cite{Cos12} repetitions in few minutes. It is therefore safe to conclude that a reasonable $p$-value can be achieved with a realistic experimental setup in a quite short realization time.

A word of caution is necessary here, since the point of the feasibility is indeed delicate and we need to stress the limitations of our proposal. The device works in a limited band, which for a JJ is roughly comprised between $100\;\text{MHz}$ and $100\;\text{GHz}$. The lower limit, i.e., $100\;\text{MHz}$, is intrinsically given by the current ramp frequency, 
as the triangular wave that drives the JJ should be chosen, even if faster electronic is available, to warranty that it is slower than a microsecond, as for instance in Ref.~\cite{Cos12}.
This limit is also instrumental in the usual assumption of an adiabatic process, which ensures that the system reaches the equilibrium at the current bias. The upper frequency limit, i.e., $100\;\text{GHz}$, is given by the typical JJ response time. In fact $\omega_J$ is typically below $1\;\text{THz}$. 
Although the circuitry in this part of the spectrum is not trivial, in this bandwidth faithful channels are available, since it is possible to couple an on-chip source, or also, even if it is more complicated, to device means to send the off-chip signal to the JJ, possibly through a transducer and a low noise channel. The faithful transmission of the off-chip noise to the JJ is beyond the scope of this paper. However, one could imagine feeding the JJ through an on-chip antenna (which hence resides at cryogenic temperature) that receives the signal to be analyzed. This technique, or similar ones, could thermally decouple the signal generator from the detector.
To summarize: the proposed method assumes that the L\'evy  noise ``signal'' power is at least comparable with, or larger than, the Gaussian noise eventually captured by the circuitry.

\section{Conclusions}
\label{Conclusions}\vskip-0.2cm
We have investigated the switching currents distributions (SCDs) in conventional Josephson junctions in the presence of a L\'evy noise source.
L\'evy distributed fluctuations are characterized by scale-free jumps or L\'evy flights. 
Consequently, we expect the SCDs to exhibit a peculiar behaviour markedly different from the Gaussian noise case. The aim is to detect the characteristics of the L\'evy noise from the SCDs.
The proposed method allows to deduce the features of L\'evy noise from the estimated CDFs of switching currents in a realistic measurement time. Moreover, the corresponding numerical simulations are viable in a reasonable time.
Specifically, depending on the value of the stability distribution index, $\alpha$, we have numerically found that: \emph{i}) for $0 < \alpha < 1$, the SCDs are peaked at zero bias current; \emph{ii}) for $\alpha \simeq 1$, the SCD is roughly flat; \emph{iii}) finally, for $1 < \alpha < 2$, the SCDs are peaked at high bias currents (alike the usual Gaussian noise induced peak) and slowly decrease at low bias currents.
A peculiar behaviour can be observed also in the cumulative distribution function (CDF) curves, that at a given value of $i_b$ decrease with increasing $\alpha$. 
Moreover, the CDFs are convex for $\alpha < 1$, and concave for $\alpha > 1$ (the case $\alpha = 1$ corresponds to a linear CDF).
 
Our results are relevant for three reasons. First, the analytic estimate, see Eq.(13), of the shape of the switching currents under a L\'evy noise source was not previously obtained and it is not a trivial exercise, for it calls for physical approximations. Second, the numerical approach shows that the approximation is good enough (in a certain range of parameters). Third, the achievements of the work pave the way towards the application of Josephson junctions to characterize L\'evy noise sources. Thus, the issue of concrete experimental estimates of the characteristic L\'evy parameters is a further, not yet explored, extension of the potentialities of Josephson-based noise detectors.

A theoretical good estimate of the SCDs can be retrieved on the basis of the Fulton adiabatic approach~\cite{Ful74} and assuming that the average escape time for the L\'evy guided overdamped case can be extended to moderately damped systems.
These theoretical findings are confirmed by the abovementioned numerical observations.
Moreover, the theoretical approach recovers a previous result~\cite{Gua17}, where a phenomenological linear approximation has been applied [see Eq.\eqref{P_t}]. 
Finally, we achieve, from the SCDs through the theoretical model [see Eqs.~\eqref{PDFib} and~\eqref{CDFib}], the estimate of the universal (i.e., barrier height independent) noise coefficient ${\cal C}_{\alpha}$ and then, if the other parameters are known, the value of the stability index $\alpha$.

Notably, we observe that the proposed method is quite robust in recognizing the L\'evy component in a noisy background, for instance, thermal. In fact, the probability of a particle to overcome a barrier when subject to L\'evy noise is independent of the barrier height. This is remarkably different from the Gaussian noise case, where the probability to overcome the barrier depends exponentially on the barrier energy. Also, if both L\'evy and Gaussian (thermal) components contribute to the overall noise level, they do not interfere, because they produce switching at different bias levels: (\emph{i}) the L\'evy noise in the lower part of the distribution at low bias currents; (\emph{ii}) the Gaussian noise, when the energy barrier becomes comparable to the noise energy, for high bias currents close to the critical value.

Finally, the Josephson-based method that we proposed offers an evident advantage when the unknown noise is characterized by fat tails, i.e., by a finite probability of a fluctuation with infinitely large intensity. This type of noise usually poses a serious difficulty to the experimentalist, for it requires extremely long times to reconstruct the behavior at large values. Thus, determining the value of the parameter $\alpha$ demands long experiments (or simulations) to explore extreme values. In contrast, sweeping the bias is very effective, because the bias increase lowers the trapping energy barrier, and therefore in a given predetermined ramp time (namely, the time it takes the linearly ramped bias current to reach the critical current) the energy barrier vanishes and a switching event is definitively recorded also for a vanishing noise intensity.

\begin{acknowledgments}
This work was supported by the Grant of the Government of the Russian Federation, contract No. 074-02-2018-330 (2). We acknowledge also partial support by Ministry of Education, University and Research of the Italian government.
C.G. acknowledge the European Research Council under the European Union's Seventh Framework Program (FP7/2007-2013)/ERC Grant agreement No.~615187-COMANCHE.
V.P. acknowledges INFN, for partial financial support through the project Virgo.
\end{acknowledgments}


%

\end{document}